\documentclass[aps,prl,reprint,groupedaddress]{revtex4-2}
\UseRawInputEncoding
\usepackage[colorlinks,linkcolor=blue,urlcolor=blue,anchorcolor=blue,citecolor=blue]{hyperref}
\usepackage{amsmath}
\usepackage{graphicx}
\usepackage{dcolumn}
\usepackage{mathrsfs}
\usepackage{amssymb}
\usepackage{amsfonts,multirow}
\usepackage{bm,changes}
\usepackage{color}
\usepackage{ulem} 
\usepackage{changes}

\begin{document}
\draft
\title{Exploring the topology induced by non-Markovian Liouvillian exceptional points}
\author{Hao-Long Zhang$^{1}$}
\thanks{These authors contribute equally to this work.}
\author{Yan Wang$^{1}$}
\thanks{These authors contribute equally to this work.}
\author{Wen Ning$^{1}$}
\thanks{These authors contribute equally to this work.}
\author{Shou-Bang Yang$^{1}$}
\author{Jia-Hao L\"{u}$^{1}$} 
\author{Fan Wu$^{1}$}
\author{Pei-Rong Han$^{2}$}
\thanks{E-mail: han\_peirong@qq.com}
\author{Li-Hua Lin$^{1,3}$} 
\author{Zhen-Biao Yang$^{1,3}$}
\thanks{E-mail: zbyang@fzu.edu.cn}
\author{Shi-Biao Zheng$^{1,3}$}
\thanks{E-mail: t96034@fzu.edu.cn}
\address{$^{1}$Fujian Key Laboratory of Quantum Information and Quantum\\
Optics, College of Physics and Information Engineering, Fuzhou University,\\
Fuzhou 350108, China\\
$^{2}$School of Physics and Mechanical and Electrical Engineering, Longyan\\
University, Longyan 364012, China\\
$^{3}$Hefei National Laboratory, Hefei 230088, China}
\date{\today }

\begin{abstract}
Non-Hermitian (NH) systems can display exotic topological phenomena without Hermitian counterparts, enabled by exceptional points (EPs). So far, investigations of NH topology have been restricted to EPs of the NH Hamiltonian, which governs the system dynamics conditional upon no quantum jumps occurring. The Liouvillian superoperator, which combines the effects of quantum jumps with NH Hamiltonian dynamics, possesses EPs (LEPs) that are significantly different from those of the corresponding NH Hamiltonian. We here study the topological features of the LEPs in the system consisting of a qubit coupled to a non-Markovian reservoir. We find that two distinct winding numbers can be simultaneously produced by executing a single closed path encircling the twofold LEP2, formed by two coinciding LEP2s, each involving a pair of coalescing eigenvectors of the extended Liouvillian superoperator. We experimentally demonstrate this purely non-Markovian phenomenon with a circuit, where a superconducting qubit is coupled to a decaying resonator which acts as a reservoir with memory effects. The results push the exploration of exceptional topology from the Markovian to non-Markovian regime.
\end{abstract}

\maketitle

\vskip0.5cm

\narrowtext

The behaviors of any quantum system are unavoidably affected by its
surrounding environment, which can be modeled as a reservoir that contains a
continuum of electromagnetic modes \cite{Scully_Zubairy_1997, PhysRevLett.101.080503}. The effects of such a reservoir can be described by a dissipator that influences the system dynamics in two ways. On the one hand, it adds a non-Hermitian (NH) term to the system Hamiltonian. On the other hand, it induces random quantum jumps, as a consequence of leakage of photons into the reservoir. When no quantum jumps occur, the
system dynamics is described by an NH Hamiltonian. The most remarkable feature of the NH Hamiltonian is the presence of exceptional points (HEPs)
\cite{NatMater.18.783,Science.363.eaar7709,Advance_in_Physics.69.249, PhysRevLett.120.146402, PhysRevX.8.031079}, where both the eigenenergies and eigenvectors coalesce. The HEPs can display exotic behaviors that are inaccessible with Hermitian systems, exemplified by the exceptional topology \cite{PhysRevLett.120.146402, PhysRevX.8.031079, PhysRevB.101.020201, RevModPhys.93.015005, NatRevPhys.4.745, PhysRevLett.86.787, Science.370.1077, SciAdv.7.eabj8905, PhysRevLett.127.090501, Science.371.1240, PhysRevLett.127.034301, Nature.598.59, PhysRevLett.130.163001, PhysRevLett.130.017201, PhysRevLett.127.186602, NatCommun.15.10293, SciBull.70.2446, Nature.607.271, Nature.537.80, Nature.537.76, PhysRevX.8.021066, PhysRevLett.126.170506}, which can manifest either in
the eigenvectors or in the complex eigenspectrum.
\begin{figure*}
	\includegraphics{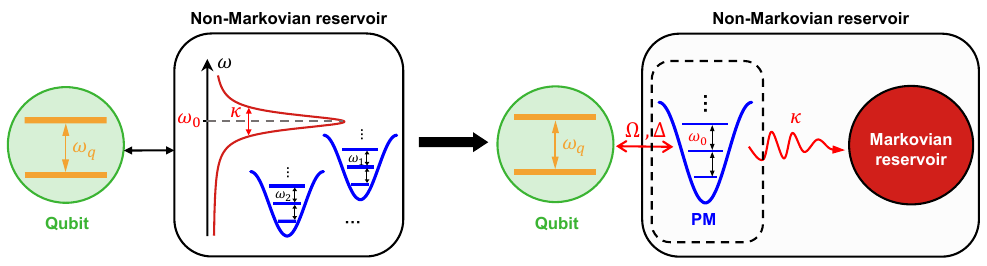} 
	\caption{(color online). Sketch of the theoretical model. A qubit with the frequency $\omega_{q}$ interacts with a non-Markovian reservoir containing a continuum of electromagnetic field modes with the spectral width $\kappa$ and central frequency $\omega_{0}$. Such a reservoir is equivalent to a leaky cavity storing an effective photonic mode, referred to as a pseudomode (PM), with the frequency $\omega_{0}$ and decaying rate $\kappa$. The coupling strength $\Omega$ and detuning $\Delta =\omega_{q}-\omega_{0}$ between the qubit and the PM serve as the control parameters of the extended Liouvillian superoperator, which governs the dynamics of the qubit-PM system.}
	\label{fig1}
\end{figure*}

Recently, it was discovered that more intriguing phenomena can emerge in the
eigenspectrum of the Liouvillian superoperator, where the quantum jump is
incorporated with the NH Hamiltonian dynamics. The corresponding
degeneracies, at which both the eigenvalues and eigenvectors of the
Liouvillian superoperator coalesce, are referred to as the Liouvillian EPs
(LEPs) \cite{PhysRevA.100.062131, PhysRevA.102.033715, PRXQuantum.2.040346, AAPPSBull.34.22}. The Liouvillian eigenspectrum can display much richer
exceptional structures than the NH Hamiltonian eigenspectrum. For example,
the Liouvillian eigenspectrum for a driven qubit in a Markovian reservoir
involves exceptional lines and higher-order EPs \cite{AAPPSBull.34.22}, which have no
counterparts in the eigenspectrum of the corresponding NH Hamiltonian.
Experimentally, Markovian LEPs have been observed in superconducting
circuits \cite{PhysRevLett.127.140504, PhysRevLett.128.110402,NewJPhys.26.123032} and ion traps \cite{NatCommun.16.7478}, and their applications in chiral state
transfer \cite{PhysRevLett.134.146602} and control of quantum heat engines \cite{NatCommun.13.6225, PhysRevLett.130.110402} have been
demonstrated.

The Markovian approximation breaks down when the reservoir is structured
\cite{RevModPhys.88.021002, ApplPhysLett.125.124003}, e.g., an electromagnetic cavity to which an atom is coupled. Such a reservoir has a memory effect, which enables backflow of the energy and information. To account for this effect, some additional degrees of freedom of the reservoir need to be introduced, which increases the dimension of the Liouvillian superoperator and its eigenvectors. The corresponding Liouvillian eigenspectrum is fundamentally different from that associated with the Markovian reservoir, featuring coexistence of LEPs with different orders as a consequence of the memory effect \cite{NatCommun.16.1289}, which has been confirmed in a very recent experiment \cite{Arxiv.2503.06977}.

These advancements naturally lead to the important question: What are the topological features of LEPs? So far, this question has not been explored in theory and experiment. We here investigate the topology associated with the
eigenspectrum of a qubit coupled to a reservoir with the Lorentzian spectral density. We discovered that the singularity in parameter space, where all LEPs are located, is characterized by a hybrid topological invariant, which
consists of different winding numbers associated with different Liouvillian
eigenenergies when the control parameters are varied along a closed path
enclosing these LEPs. We experimentally implement this unique topological
feature with a circuit quantum electrodynamics system, where a Josephson-junction-based superconducting qubit is controllably coupled to
its readout resonator that serves as a non-Markovian reservoir for the
qubit. The coupling strength and detuning span a parameter space, where a
desired cyclic path is executed by controlling the amplitude and frequency
of the ac flux that mediates a sideband interaction between the qubit and
the resonator.

The theoretical model consists of a qubit coupled to a reservoir composed of
a continuum of electromagnetic field modes with a Lorentzian spectral
density \cite{Scully_Zubairy_1997}%
\begin{equation}
J(\omega )=\frac{1}{\pi }\frac{\kappa /2}{(\omega -\omega _{0})^{2}+(\kappa
/2)^{2}},
\end{equation}%
where $\kappa $ represents the spectral width and $\omega _{0}$ is the
spectral central frequency. The memory effect of this reservoir can be
captured by introducing a pseudomode (PM) with the frequency $\omega _{0}$
and the decaying rate $\kappa $. In other words, the non-Markovian reservoir
corresponds to a leaky cavity holding a photonic mode with the frequency $%
\omega _{0}$, as shown in Fig.~\ref{fig1}. In the framework rotating at the frequency 
$\omega _{0}$, the evolution of the composite qubit-PM system is described by the master equation
\begin{equation}
\frac{d\rho }{dt}={\cal L}\rho ,
\end{equation}%
where ${\cal L}$ is the extended Liouvillian operator, defined as%
\begin{equation}\label{Liou}
{\cal L}\rho =i\left[\rho H_{NH}^{\dagger }-H_{NH}\rho \right]+\kappa b\rho b^{\dagger
}.
\end{equation}%
Here $H_{NH}$ is the NH Hamiltonian, given by ($\hbar = 1$ is set)
\begin{equation}
H_{NH}=\Omega \left(b^{\dagger }\left\vert g\right\rangle \left\langle
e\right\vert +b\left\vert e\right\rangle \left\langle g\right\vert \right)+\Delta
\left\vert e\right\rangle \left\langle e\right\vert -\frac{i\kappa }{2}%
b^{\dagger }b,
\end{equation}%
where $\Omega $ and $\Delta $ are the coupling strength and detuning between the qubit and the PM, $\left\vert g\right\rangle$ and $\left\vert e\right\rangle$ represent the lower and upper levels of the qubit, and $b^{\dagger}$ and $b$ denote the creation and annihilation operators for the
PM, respectively.
\begin{figure*}
	\includegraphics{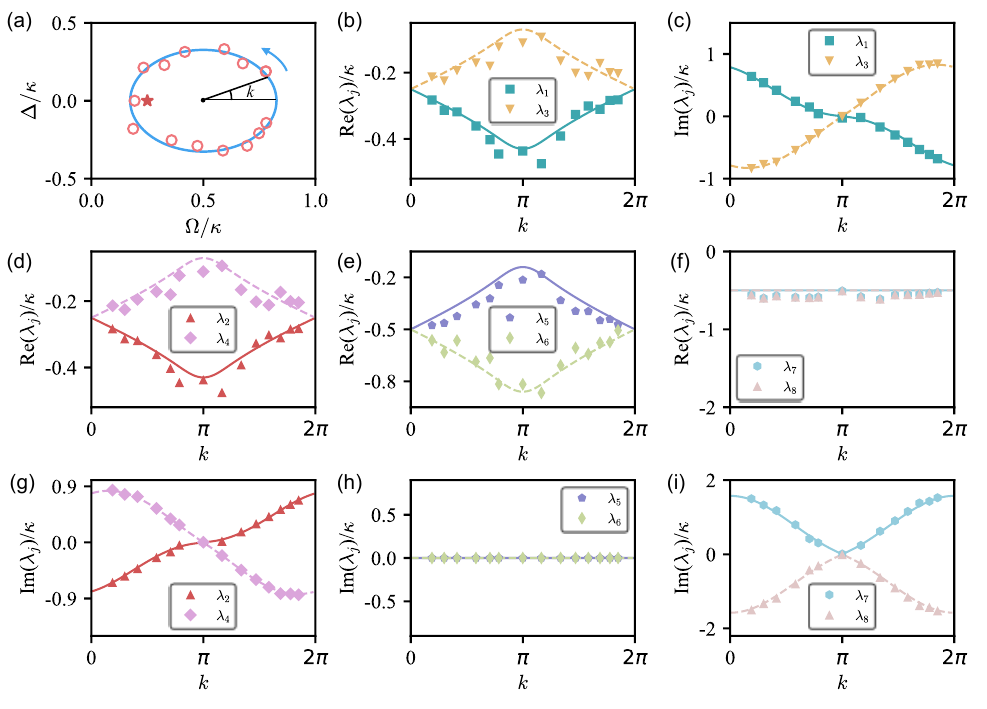} 
	\caption{(color online). Experimental measurement of the Liouvillian spectrum. (a) Parameter-space loop. We here set $\Omega =\kappa /2+r\cos k$ and $\Delta=r\sin k$, so that the LEPs (star), located at $\left\{ r=\kappa/4,k=\pi \right\} $, are enclosed in the loop when $r>\kappa /4$. (b)-(i) Measured Liouvillian eigenvalues, $\lambda_{1}$ to $\lambda_{8}$, versus $k$. $\lambda_{j}$ for each value of $k$ is extracted by tracking the dynamical evolution of the qubit-PM system starting with the initial state $(\left\vert g\right\rangle +i\left\vert e\right\rangle )\left\vert 0\right\rangle /\sqrt{2}$, and exponentially fitting the amplitude associated with the eigenvector ${\bf V}_{j}$, $A_{j}(t)=A_{j}(0)e^{\lambda_{j}t}$. All the data are measured for $r\approx0.327\kappa$.}
	\label{fig3}
\end{figure*}

The system excitation, defined as $N_{e}=\left\vert e\right\rangle
\left\langle e\right\vert +b^{\dagger }b$, is conserved under the NH
Hamiltonian. The last term of ${\cal L}\rho $ describes the quantum jump,
which induces a jump for the PM from the n-photon Fock state $\left\vert
n\right\rangle $ to $\left\vert n-1\right\rangle $. When the qubit is
initially in a superposition of $\left\vert g\right\rangle $ and $\left\vert
e\right\rangle $ and the PM initially in the vacuum state $\left\vert
0\right\rangle $, the qubit-PM evolution is restricted in the subspace with zero and one excitation: \{$\left\vert g,0\right\rangle ,\left\vert
e,0\right\rangle ,\left\vert e,1\right\rangle $\}, where the extended
Liouvillian superoperator corresponds to a $9\times 9$ NH matrix
\cite{NatCommun.16.1289, Arxiv.2503.06977}, ${\cal L}_{matrix}$, whose eigenvalues and eigenvectors satisfy the
secular equation%
\begin{equation}
{\cal L}_{matrix}{\bf V}_{j}=\lambda _{j}{\bf V}_{j}\text{,}
\end{equation}%
where ${\bf V}_{j}$ is the eigenvector with the eigenvalue $\lambda _{j}$. $%
{\cal L}_{matrix}$ possesses 9 eigenvalues given by
\begin{eqnarray}\label{eigen}
\lambda _{0} &=&0,  \nonumber \\
\lambda _{1(2)} &=&\eta _{\mp}-\sqrt{\eta _{\pm}^{2}-\Omega ^{2}},  \nonumber \\
\lambda _{3(4)} &=&\eta _{\mp}+\sqrt{\eta _{\pm}^{2}-\Omega ^{2}},  \nonumber \\
\lambda _{5(6)} &=&-\frac{\kappa }{2}\pm\sqrt{\alpha +\beta },  \nonumber \\
\lambda _{7(8)} &=&-\frac{\kappa }{2}\pm\sqrt{\alpha -\beta },  
\end{eqnarray}
where $\eta_{\pm }=(-\kappa \pm 2i\Delta )/4$, $\alpha =\kappa
^{2}/8-\Delta ^{2}/2-2\Omega ^{2}$, and $\beta =\sqrt{\alpha ^{2}+\Delta
^{2}\kappa ^{2}/4}$.

The last eight eigenvalues versus the coupling strength $\Omega$ and
detuning $\Delta$ are displayed in the Supplemental Material.
For $\Delta =0$, there exist three
pairs of degenerate eigenvectors ${\bf V}_{1}$-${\bf V}_{2}$, ${\bf V}_{3}$-$%
{\bf V}_{4}$, and ${\bf V}_{7}$-${\bf V}_{8}$ (${\bf V}_{5}$-${\bf V}_{6}$), irrespective of the value of $%
\Omega $. When $\Omega<\kappa/4$, ${\bf V}_{7}$ and ${\bf V}_{8}$ degenerate, otherwise ${\bf V}_{5}$ and ${\bf V}_{6}$ degenerate. At the parameter-space position $\left\{ \Omega =\kappa /4;\Delta
=0\right\} $, ${\bf V}_{1}$ (${\bf V}_{2}$) and ${\bf V}_{3}$ (${\bf V}_{4}$%
) coalesce, and $\lambda_{1}$ ($\lambda_{2}$) coincides with $\lambda_{3}$
($\lambda_{4}$). In addition to featuring a twofold order-2 LEP (LEP2),
this position also corresponds to an order-3 LEP (LEP3), characterized by
the coalescence of three degenerate eigenvectors ${\bf V}_{5}$, ${\bf V}_{6}$%
, and ${\bf V}_{7}$. These LEPs have been observed by a previous circuit quantum electrodynamics
experiment \cite{Arxiv.2503.06977}, but where $\Delta$ was fixed to $0$ and the topological
features of these LEPs were not investigated. As shown in Eq.~(\ref{eigen}),
when $\Delta \neq 0$, the degeneracies of ${\bf V}_{1}$-${\bf V}_{2}$, ${\bf %
V}_{3}$-${\bf V}_{4}$, and ${\bf V}_{7}$-${\bf V}_{8}$ pairs are lifted. The
variation of both the parameters $\Omega$ and $\Delta$ along a cyclic path
is necessary to probe the topology of the EPs.

The superconducting circuit used to probe the exceptional topology consists
of a bus resonator (R$_{b}$) connecting five frequency-tunable qubits (Q$_{j}
$ with $j=1$ to $5$), each of which is connected to a readout resonator (R$_{j}$). In the experiment, Q$_{1}$ is used as the test qubit for
constructing the LEPs, and R$_{1}$ serves as the non-Markovian reservoir,
which has a Lorentzian spectral distribution with a central frequency $%
\omega_{0}/2\pi\approx6.66$ MHz and a spectral width $\kappa \approx 5$ $\mu s^{-1}$. The maximum frequency of Q$_{1}$ is $\omega_{\max }/2\pi\approx6.05$ MHz. The difference between $\omega_{0}$ and $%
\omega_{\max }$ is much larger than the on-resonance coupling strength
between Q$_{1}$ and the decaying photonic mode of R$_{1}$ (PM), which has a
value of $g_{r}\approx 2\pi \times 40  $ MHz. The controllable Q$_{1}$-PM coupling is realized with an ac flux, which modulates the frequency of the qubit as \cite{Arxiv.2503.06977, NatCommun.12.5924,PhysRevLett.131.260201}, $\omega_{q}=%
\stackrel{-}{\omega }+\varepsilon \cos (\nu t)$, where $\stackrel{-}{\omega }
$ is the mean frequency of the qubit, and $\varepsilon $ and $\nu $
denote the modulating amplitude and frequency, respectively. When $\nu $ is
close to $\omega_{0}-\stackrel{-}{\omega }$, Q$_{1}$ interacts with R$_{1}$
at the first sideband with respect to the parametric modulation. The
effective photon swapping rate between Q$_{1}$ and R$_{1}$ is $\Omega
=g_{r}J_{1}(\mu )$, and the detuning is $\Delta =\stackrel{-}{\omega }+\nu
-\omega _{0}$, where $J_{1}(\mu )$ is the first-order Bessel function of the
first kind with $\mu =\varepsilon /\nu $. The interplay between this
coherent coupling and the incoherent photonic decay of R$_{1}$ is described
by the extended Liouvillian operator ${\cal L}$ of Eq.~(\ref{Liou}). The second qubit
Q$_{2}$ is used as an ancilla for the readout of the output Q$_{1}$-PM
state. The other three qubits are not used and are tuned to effectively decouple from Q$_{1}$ and Q$_{2}$ throughout the experiment due to the large detunings.
\begin{figure}
	\includegraphics[width=1.0\linewidth]{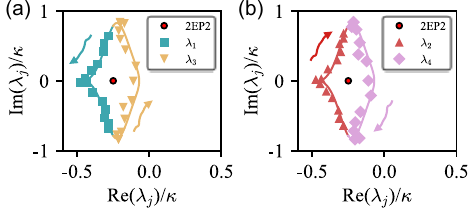} 
	\caption{(color online). Characterization of the hybrid topological invariant. Trajectories of the measured Liouvillian eigenvalues: $\lambda_{1}$ and $\lambda_{3}$ (a); $\lambda_{2}$ and $\lambda_{4}$ (b);  are plotted on the complex $\mathop{\rm Re}(\lambda) $-$\mathop{\rm Im}(\lambda)$ plane (in unit of $\kappa$). The control parameters are varied along the circle shown in Fig.~\ref{fig3}. The solid lines denote the trajectories of the calculated eigenvalues.}
	\label{fig4}
\end{figure}

\bigskip In our experiment, the Liouvillian spectrum is extracted from the
dynamically evolved output state, which can be expressed in terms of the
eigenvectors of the Liouvillian superoperator
\begin{equation}
{\bf V}(t){\bf =}\sum_{j}A_{j}(t){\bf V}_{j},
\end{equation}%
where $A_{j}(t)=A_{j}(0)e^{\lambda_{j}t}$. The eigenvalue $\lambda_{j}$ is
inferred by exponentially fitting the evolution of the amplitude $A_{j}(t)$.
When $A_{j}(0)=0$, ${\bf V}_{j}$ does not appear in the expansion of ${\bf V}%
(t)$ so that $\lambda _{j}$ cannot be obtained from the dynamical evolution
of the associated amplitude $A_{j}(t)$, which remains zero. Therefore, the
number of eigenvalues that can be extracted depends upon the initial
state. In our experiment, we choose $(\left\vert g\right\rangle +i\left\vert
e\right\rangle )/\sqrt{2}$ as the initial state for Q$_{1}$, which can be
prepared from the ground state with a $\pi /2$ pulse. The resonator R$_{1}$
is initially in the vacuum state $\left\vert 0\right\rangle $. This choice
ensures that each of the eigenvectors has a nonvanishing amplitude
during the state evolution for $\Delta\neq0$. After preparation of this initial state, the
parametric modulation is applied to mediate the sideband interaction between
Q$_{1}$ and the PM. The effective interaction strength $\Omega $ and
detuning $\Delta $ are controlled by the modulating parameters $\varepsilon $
and $\nu $, respectively.

Following the switch-off of the parametric modulation, it is necessary to perform the state mappings: Q$_{1}\rightarrow $R$_{b}$, R$_{b}\rightarrow $Q$_{2}$, and PM$\rightarrow $Q$_{1}$. These operations transfer the output PM-Q$_{1}$ state to the Q$_{1}$-Q$_{2}$ system, whose joint density matrix can be reconstructed by quantum state tomography. From the measured output density matrix, we can extract the amplitude $A_{j}(t)$ associated with the eigenvector ${\bf V}_{j}$ of the extended Liouvillian superoperator. The corresponding eigenvalue $\lambda_{j}$ is inferred by exponentially fitting the data of $A_{j}$ measured for different Q$_{1}$-PM interaction times.

To reveal the topology of the complex eigenspectrum, it is necessary to map out the corresponding eigenenergies along a circuit in the parameter space $\left\{ \Omega ,\Delta \right\} $. We choose circular loops centered at $\left\{ \Omega =\kappa /2;\Delta =0\right\} $. With this setting, we can rewrite the two control parameters as $\Omega =\kappa /2+r\cos k$ and $\Delta /2=r\sin k$, and the LEPs are enclosed in the loop when $r>\kappa /4$. In our experiment, the topology is demonstrated with the choice $r=0.327\kappa$, as shown in Fig.~\ref{fig3}(a). Fig.~\ref{fig3}(b)-(i) display the eigenvalues $\lambda_{j}$ for $j=1$ to $8$, inferred by exponential fitting of $A_{j}(t)$ for different values of $k$ (see Supplemental Material for details). The eigenvalue $\lambda_{j}$ is inferred by the fitted amplitude associated with the eigenvector ${\bf V}_{j}$. 

The spectral topology is manifested on the complex $%
\mathop{\rm Re}%
\lambda $-$%
\mathop{\rm Im}%
\lambda $ plane. The winding number associated with $\lambda_{j}$ is
defined as \cite{PhysRevB.101.020201, PhysRevLett.130.163001} 
\begin{equation}\label{wind}
{\cal W}_{j}=\frac{1}{2m\pi }\int_{0}^{2m\pi }dk\partial_{k}\arg [\lambda_{j}(k)-\lambda_{\rm EP}],
\end{equation}%
where $m$ is the smallest positive integer that satisfies the condition $%
\lambda_{j}(k+2m\pi )=\lambda_{j}(k)$. The absolute value of ${\cal W}_{j}$
counts the number of times that the eigenvalue $\lambda_{j}$ winds around
the EP on the complex plane, and its sign denotes the moving direction of $%
\lambda_{j}$ when $k$ is varied from $0$ to $2m\pi $. Fig.~\ref{fig4}(a) presents the
measured eigenvalues $\lambda_{1}$ and $\lambda_{3}$ on the complex plane,
which well agree with the ideal results (solid lines). As expected, when $k$
is varied from 0 to $2\pi $, the combination of the paths traversed by $%
\lambda_{1}$ and $\lambda_{3}$ forms a loop enclosing the degenerate
eigenvalue $\lambda_{\rm LEP2}=-\kappa /4$ at the twofold LEP2. At $k=2\pi $,
an exchange of eigenvalues is completed, that is, $\lambda _{1}$ ($\lambda
_{3}$) is moved to the original position of $\lambda _{3}$ ($\lambda _{1}$).
As the variation of the control parameters for $k=2\pi$ to $4\pi $ is
identical to that for $k=0$ to $2\pi $, the trajectory traversed by $\lambda
_{3}$ ($\lambda _{1}$) during $k=0$ to $2\pi $ corresponds to that by $%
\lambda _{1}$ ($\lambda _{3}$) during the interval from $k=2\pi $ to $4\pi $%
. In other words, during $k=0$ to $4\pi $ these two eigenvalues are
exchanged for two times, traversing a closed path that encloses the LEP2.
This implies that the corresponding winding numbers are ${\cal W}_{1}={\cal W%
}_{3}=1/2$. Fig.~\ref{fig4}(b) displays the measured eigenvalues $\lambda_{2}$ and $%
\lambda_{4}$ on the $%
\mathop{\rm Re}%
\lambda $-$%
\mathop{\rm Im}%
\lambda $ plane. The trajectories traversed by $\lambda _{2}$ and $\lambda_{4}$ also form a loop that has the same shape but opposite direction as compared with that formed by the paths of $\lambda _{1}$ and $\lambda _{3}$. Consequently, ${\cal W}_{2}={\cal W}_{4}=-1/2$. Such a hybrid winding number is a unique feature of the twofold LEP2, which is formed by two overlapping LEP2s, one characterizing the coalescence of the eigenvectors ${\bf V}_{1}$ and ${\bf V}_{3}$, while the other featuring the coincidence between ${\bf V}_{2}$ and ${\bf V}_{4}$. 

The most remarkable topological feature of the system is that the winding numbers associated with trajectories traversed by $\lambda _{1(3)}$ and $\lambda _{2(4)}$ have the opposite values. This implies that the twofold LEP2 has a hybrid topological invariant, which can be either $1/2$ or $-1/2$, depending on which pair of eigenvalues is used to extract the winding number. This exotic feature distinguishes the topological phenomena of the non-Markovian LEP2 from those of the corresponding HEP2, which has a definite winding number. This hybrid topological invariant is fundamentally different from that possessed by the exceptional nexus \cite{Science.370.1077}, which manifests as distinct winding numbers associated with Berry phases accumulated by cyclic paths on different complex planes of the control parameters. In the present system, opposite winding numbers are produced by the same cyclic path in parameter space. Such an intriguing exceptional topological phenomenon is a purely non-Markovian effect, which emerges only when the information transferred to the PM of the reservoir can flow back to the qubit and be extracted. Unlike the LEP2, the LEP3 does not exhibit any topological feature, as shown in the Supplemental Material.

In conclusion, we have explored the topological phenomena of Liouvillian EPs of a qubit coupled to a non-Markovian reservoir in a superconducting circuit. The topology is encoded in the Liouvillian spectrum in parameter space, which is mapped out by fitting the evolutions of the amplitudes associated with the eigenvectors of the extended Liouvillian superoperator. The twofold LEP2, which is a consequence of the non-Markovianity, possesses a hybrid topological invariant, manifested by opposite winding numbers associated with the two LEP2s that coincide with each other both in the parameter space and on the complex spectrum plane, but carry opposite topological charges. Our work paves the way for the exploration of exotic topological phenomena associated with non-Markovian quantum EPs.

\textit{Acknowledgments}---This work was supported by the National Natural Science Foundation of China (Grants No. 12474356, No. 12475015, No. 12274080, No. 12505021, No. 12505016), Quantum Science and Technology-National Science and Technology Major Project (Grant No. 2021ZD0300200), and the Natural Science Foundation of Fujian Province (Grants No. 2025J01383, No. 2025J01465).

\textit{Data availability}---The data that support the findings of this article are not publicly available. The data are available from the authors upon reasonable request.

\bibliography{reference}

\end{document}


\title{Supplementary Materials for \\``Exploring the topology induced by non-Markovian Liouvillian exceptional points"}
	
\author{Hao-Long Zhang$^{1}$}
\thanks{These authors contribute equally to this work.}
\author{Yan Wang$^{1}$}
\thanks{These authors contribute equally to this work.}
\author{Wen Ning$^{1}$}
\thanks{These authors contribute equally to this work.}
\author{Shou-Bang Yang$^{1}$}
\author{Jia-Hao L\"{u}$^{1}$} 
\author{Fan Wu$^{1}$}
\author{Pei-Rong Han$^{2}$}
\thanks{E-mail: han\_peirong@qq.com}
\author{Li-Hua Lin$^{1,3}$} 
\author{Zhen-Biao Yang$^{1,3}$}
\thanks{E-mail: zbyang@fzu.edu.cn}
\author{Shi-Biao Zheng$^{1,3}$}
\thanks{E-mail: t96034@fzu.edu.cn}
\address{$^{1}$Fujian Key Laboratory of Quantum Information and Quantum\\
Optics, College of Physics and Information Engineering, Fuzhou University,\\
Fuzhou 350108, China\\
$^{2}$School of Physics and Mechanical and Electrical Engineering, Longyan\\
University, Longyan 364012, China\\
$^{3}$Hefei National Laboratory, Hefei 230088, China}

\vskip0.5cm

\narrowtext

\maketitle
	
\tableofcontents
\section{Extended Liouvillian matrix and solution of the secular equation}
The dynamics of the entire Q-PM system can be described by the Lindblad master equation
\begin{equation}
\frac{\partial\rho}{\partial t} = -i [H,\rho] + \kappa b \rho b^{\dagger}-\frac{\kappa}{2}\{b^{\dagger}b,\rho\} \equiv \mathcal{L} \rho,
\end{equation}
where the system Hamiltonian is given by $H=g(b^{\dagger}|g\rangle\langle e|+b|e\rangle\langle g|)+\Delta\vert e\rangle\langle e\vert$, the system state $\rho$ is a $3\times3$ density matrix in the subspace $\{|g,0\rangle,|e,0\rangle,|g,1\rangle\}$, and the extended Liouvillian superoperator $\mathcal{L}_{matrix}$ in the defined matrix representation can be written as
\begin{equation}
\begin{aligned}
\mathcal{L}_{matrix} &= -i\left(H \otimes I - I \otimes H^{\top}\right) 
+ \frac{\kappa}{2} \left(2 b \otimes b^{*} -b^{\dagger}b \otimes I - I \otimes b^{\top}b^{*}\right)\\
&= -i\left(H \otimes I - I \otimes H^{\top}\right) 
+ \frac{\kappa}{2} \left(2 b \otimes b -b^{\dagger}b \otimes I - I \otimes b^{\dagger}b\right).
\end{aligned}
\end{equation}
Here $\otimes$ is the Kronecker product operation, $\top$ and $*$ represent the transpose and complex conjugate operations, respectively. Therefore, in the subspace spanned by $\{|g,0\rangle,|e,0\rangle,|g,1\rangle\}$, the matrix form of the extended Liouvillian superoperator is a $9 \times 9$ matrix:
\begin{equation}
\mathcal{L}_{matrix} =\left(
\begin{array}{ccccccccc}
0 & 0 & 0 & 0 & 0 & 0 & 0 & 0 & \kappa \\
0 & i\Delta & ig & 0 & 0 & 0 & 0 & 0 & 0 \\
0 & ig & -\frac{\kappa}{2} & 0 & 0 & 0 & 0 & 0 & 0 \\
0 & 0 & 0 & -i\Delta & 0 & 0 & ig & 0 & 0 \\
0 & 0 & 0 & 0 & 0 & ig & 0 & -ig & 0 \\
0 & 0 & 0 & 0 & ig & -\frac{\kappa}{2}-i\Delta & 0 & 0 & -ig \\
0 & 0 & 0 & -ig & 0 & 0 & -\frac{\kappa}{2} & 0 & 0 \\
0 & 0 & 0 & 0 & -ig & 0 & 0 & -\frac{\kappa}{2}+i\Delta & ig \\
0 & 0 & 0 & 0 & 0 & -ig & 0 & ig & -\kappa \\

\end{array}
\right).
\end{equation}
By solving the secular equation
\begin{figure*}
	\includegraphics{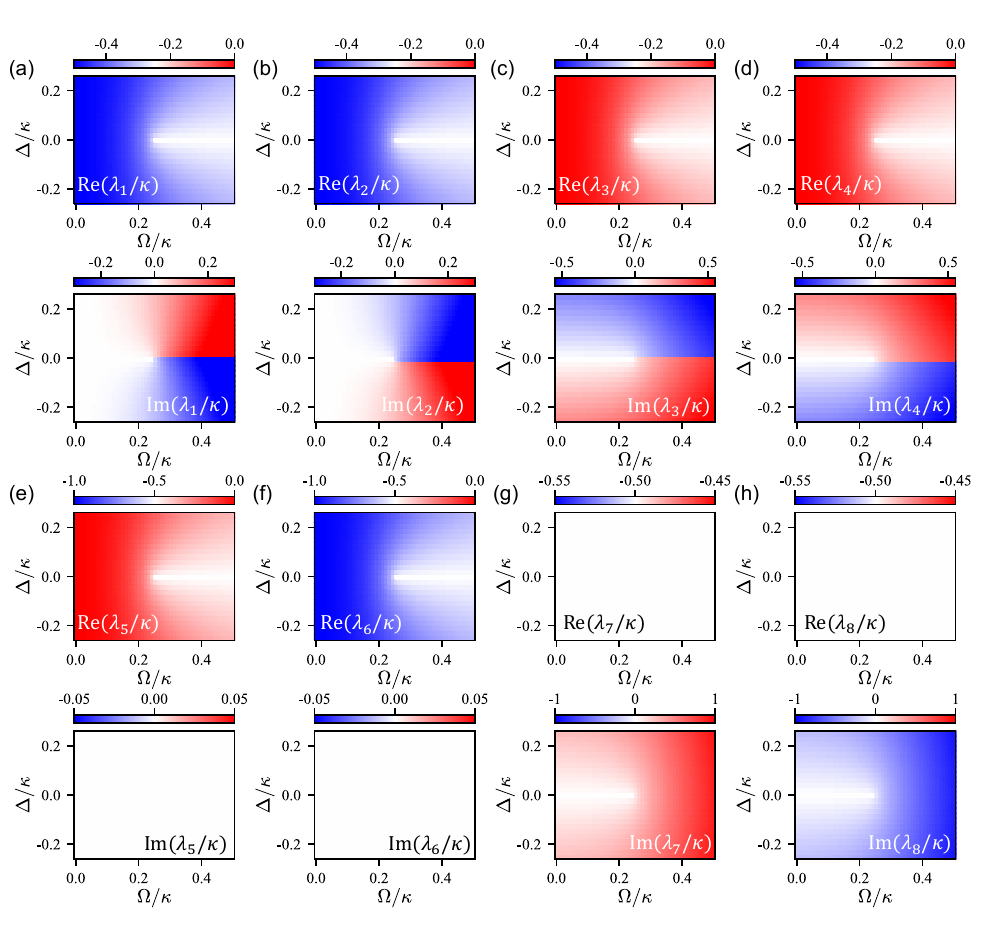} 
	\caption{(color online). Liouvillian spectrum for the theoretical model. (a)-(h) The last eight eigenvalues of the extended Liouvillian superoperator, $\lambda_{1}$ to $\lambda_{8}$, versus the control parameters $\Omega$ and $\Delta$. The first eigenvalue $\lambda_{0}$ remains to be $0$, irrespective of $\Omega$ and $\Delta$ (in unit of $\kappa$).}
	\label{fig_spectrum}
\end{figure*}

\begin{equation}
\mathcal{L}_{matrix} \mathbf{V}_j = \lambda_j \mathbf{V}_j,
\end{equation}
the eigenvalues $\lambda_j$ and the eigenvectors $\mathbf{V}_j$ (represented by the density matrix form $\rho_j$) of the extended Liouvillian superoperator can be obtained, written as follows: 
\begin{equation}
\lambda_0 = 0,\quad
\mathbf{V}_0 =
\left(
\begin{array}{cccccccccc}
1 & 0 & 0 &
0 & 0 & 0 &
0 & 0 & 0 
\end{array}
\right)^{\rm T};
\end{equation}

\begin{equation}
\lambda_1  = \eta_{-}-\sqrt{\eta_+^2-\Omega^2}, \quad
\mathbf{V}_1 =\frac{1}{N_1}
\left[
\begin{array}{cccccccccc}
0 & 0 & 0 &
-i\left(\eta_{+}+\sqrt{\eta_+^2-\Omega^2}\right)  & 0 & 0 &
\Omega & 0 & 0
\end{array}
\right]^{\rm T};
\end{equation}

\begin{equation}
\lambda_2  = \eta_{+}-\sqrt{\eta_-^2-\Omega^2}, \quad
\mathbf{V}_2 =\frac{1}{N_2}
\left[
\begin{array}{cccccccccc}
0 & i\left(\eta_{-}+\sqrt{\eta^2_{-}-\Omega^2} \right) & \Omega &
0  & 0 & 0 &
0 & 0 & 0
\end{array}
\right]^{\rm T};
\end{equation}

\begin{equation}
\lambda_3  = \eta_{-}+\sqrt{\eta_+^2-\Omega^2}, \quad
\mathbf{V}_3 =\frac{1}{N_3}
\left[
\begin{array}{cccccccccc}
0 & 0 & 0 &
-i\left(\eta_{+}-\sqrt{\eta_+^2-\Omega^2}\right)  & 0 & 0 &
\Omega & 0 & 0
\end{array}
\right]^{\rm T};
\end{equation}

\begin{equation}
\lambda_4  = \eta_{+}+\sqrt{\eta_-^2-\Omega^2}, \quad
\mathbf{V}_4 =\frac{1}{N_4}
\left[
\begin{array}{cccccccccc}
0 & i\left(\eta_{-}-\sqrt{\eta^2_{-}-\Omega^2} \right) & \Omega &
0  & 0 & 0 &
0 & 0 & 0
\end{array}
\right]^{\rm T};
\end{equation}

\begin{equation}
\lambda_5  = -\frac{\kappa}{2}+\sqrt{\alpha+\beta}, \quad
\mathbf{V}_5 =\frac{1}{N_5}
\left[
\begin{array}{cccccccccc}
 \frac{\kappa}{\lambda_5}&\\ 0 &\\ 0 &\\
0  &\\ \frac{1}{16\Omega^2}\left(\kappa^2+8\beta+4\kappa\sqrt{\alpha+\beta}+\Delta^2\left(\frac{4\sqrt{\alpha+\beta}+4\kappa}{\sqrt{\alpha+\beta}}\right) \right) &\\ \frac{i\sqrt{2}(-16\Omega^2+8\beta)+(2\Delta+i\kappa)(-2i\sqrt{2}\Delta+\sqrt{2}\kappa+4\sqrt{2}\sqrt{\alpha+\beta})}{16\sqrt{2}\Omega\sqrt{\alpha+\beta}} &\\
0 &\\ \frac{1}{8\Omega}\left(\Delta\left(\frac{4\sqrt{\alpha+\beta}+2\kappa}{\sqrt{\alpha+\beta}}-i\left(2\kappa+4\sqrt{\alpha+\beta} \right) \right) \right) &\\ 1
\end{array}
\right];
\end{equation}

\begin{equation}
\lambda_6  = -\frac{\kappa}{2}-\sqrt{\alpha+\beta}, \quad
\mathbf{V}_6 =\frac{1}{N_6}
\left[
\begin{array}{cccccccccc}
 \frac{\kappa}{\lambda_6}&\\ 0 &\\ 0 &\\
0  &\\ \frac{1}{16\Omega^2}\left(\kappa^2+8\beta-4\kappa\sqrt{\alpha+\beta}+\Delta^2\left(\frac{4\sqrt{\alpha+\beta}-4\kappa}{\sqrt{\alpha+\beta}}\right) \right) &\\ \frac{1}{8\Omega}\left(4\Delta+2i\kappa-\frac{2\Delta\kappa}{\sqrt{\alpha+\beta}}-4i\sqrt{\alpha+\beta}\right) &\\
0 &\\ \frac{1}{8\Omega}\left(\Delta\left(\frac{4\sqrt{\alpha+\beta}-2\kappa}{\sqrt{\alpha+\beta}}-i\left(-2\kappa+4\sqrt{\alpha+\beta} \right) \right) \right) &\\ 1
\end{array}
\right];
\end{equation}

\begin{equation}
\lambda_7  = -\frac{\kappa}{2}+\sqrt{\alpha-\beta}, \quad
\mathbf{V}_7 =\frac{1}{N_7}
\left[
\begin{array}{cccccccccc}
 \frac{\kappa}{\lambda_7}&\\ 0 &\\ 0 &\\
0  &\\ \frac{1}{16\Omega^2}\left(\kappa^2-8\beta-4\kappa\sqrt{\alpha-\beta}+\Delta^2\left(\frac{4i\sqrt{\alpha-\beta}+4\kappa}{\sqrt{\alpha-\beta}}\right) \right) &\\ \frac{i\sqrt{2}(16\Omega^2+8\beta)+(2\Delta+i\kappa)(-2i\sqrt{2}\Delta+\sqrt{2}\kappa+4\sqrt{2}\sqrt{\alpha-\beta})}{16\sqrt{2}\Omega\sqrt{\alpha-\beta}} &\\
0 &\\ \frac{1}{8\Omega}\left(\Delta\left(\frac{4\sqrt{\alpha-\beta}-2\kappa}{\sqrt{\alpha-\beta}}+i\left(-2\kappa+4\sqrt{\alpha-\beta} \right) \right) \right) &\\ 1
\end{array}
\right];
\end{equation}

\begin{equation}
\lambda_8  = -\frac{\kappa}{2}-\sqrt{\alpha-\beta}, \quad
\mathbf{V}_8 =\frac{1}{N_8}
\left[
\begin{array}{cccccccccc}
 \frac{\kappa}{\lambda_8}&\\ 0 &\\ 0 &\\
0  &\\ \frac{1}{16\Omega^2}\left(\kappa^2-8\beta+4\kappa\sqrt{\alpha-\beta}+\Delta^2\left(\frac{4\sqrt{\alpha-\beta}+4\kappa}{\sqrt{\alpha-\beta}}\right) \right) &\\ \frac{1}{8\Omega}\left(4\Delta+2i\kappa-\frac{2\Delta\kappa}{\sqrt{\alpha-\beta}}-4i\sqrt{\alpha+\beta}\right) &\\
0 &\\ \frac{1}{8\Omega}\left(\Delta\left(\frac{4\sqrt{\alpha-\beta}+2\kappa}{\sqrt{\alpha-\beta}}-i\left(2\kappa+4\sqrt{\alpha-\beta} \right) \right) \right) &\\ 1
\end{array}
\right].
\end{equation}
Here $\eta_{\pm} = \frac{1}{4}\left(-\kappa\pm2i\Delta\right)$,  $\alpha = \kappa^2/8-\Delta^2/3-2\Omega^2$, $\beta=\sqrt{\alpha^2+\Delta^2\kappa^2/4}$ and $N_j$ ($j=1,2,...,8$) are the normalization constants.

The last eight eigenvalues versus the coupling strength $\Omega$ and detuning $\Delta$ are displayed in Fig.~\ref{fig_spectrum}(a)-(h), where the upper and lower panels in each subfigure respectively denote the real and imaginary parts of the corresponding eigenvalue.
\begin{figure*}
    \centering
    \includegraphics[width=1.0\textwidth]{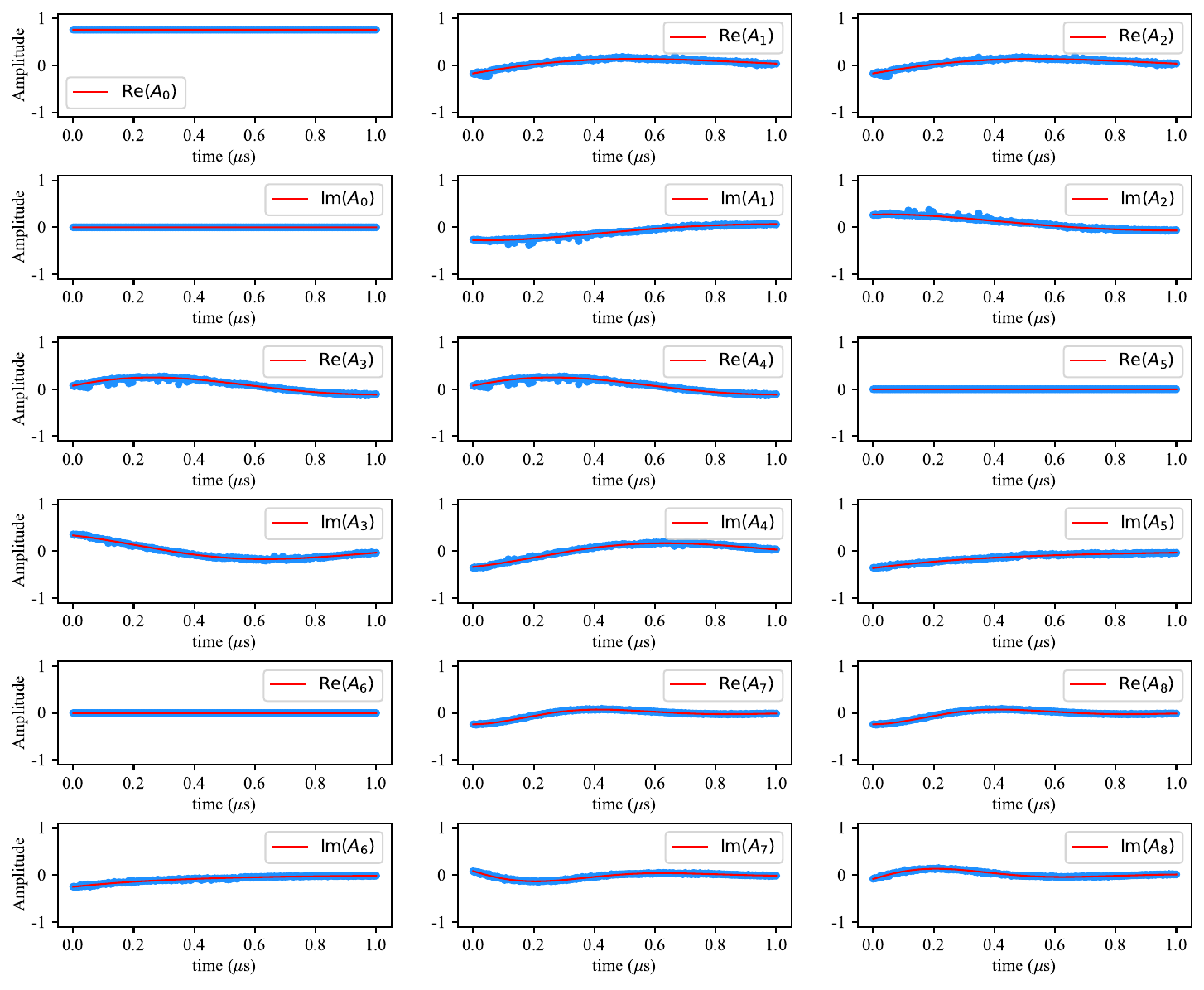}
	\caption{The real and imaginary parts of the amplitudes associated with the eigenmatrices $A_j(t)$ versus the evolution time $t$, for $g/2\pi \approx 0.620$ MHz and $\Delta/2\pi \approx 0.150$ MHz. Dots and lines denote the experimentally measured and fitting results, respectively.}
    \label{fig1}
\end{figure*}

\section{Eigenvalues extraction}

The eigenvalues of the Liouvillian superoperator can be extracted by fitting the time evolution of the amplitude $A_j(t)$, as shown in Eq.~(8) in the main text. In our experiment, the system state $\rho(t)$ at each moment can be reconstructed via quantum state tomography and expressed in terms of the superoperator ${\bf V}(t)$. To obtain the amplitude $A_j(t)$, we numerically compute the left eigenmatrices (${\bf T}_j$) of the Liouvillian superoperator ($\mathcal{L}_{matrix}$) by solving 
\begin{equation}
    \mathcal{L}_{matrix}^{\dagger}{\bf T}_j=\lambda_j^* {\bf T}_j.
\end{equation}
The amplitude $A_j(t)$ is then extracted by left-multiplying ${\bf V}(t)$ with ${\bf T}^{\dagger}_j$, according to the biorthogonal condition ${\bf T}_i^{\dagger}{\bf V}_j=\delta_{ij}$.  Through identifying the information of all the components of the density matrix $\rho(t)$ at different times, and then with resorting to the least-square method for using the obtained $A_j(t)$ to fit the exponential function $C_j e^{B t}$, we can get the best-fit eigenvalues, i.e., $B = \lambda_j$.

Under the circumstances, the error function is defined as \cite{Nat.Commun.15.10293, Arxiv.2503.06977}
\begin{equation}\label{erf}
\text{Erf} = \vert A_j(t) - C e^{Bt}\vert^2.
\end{equation}

The best-fit eigenvalues $\lambda_j$ are then the values minimizing the error function of (\ref{erf}). The experimentally measured $A_j(t)$ (dots) and their fitting results (lines) are given in Fig.~\ref{fig1}.

\begin{figure*}
	\includegraphics[width=0.5\linewidth]{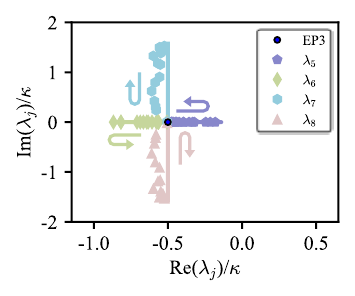} 
	\caption{(color online). Trajectories of the measured Liouvillian eigenvalues: $\lambda_{5}$, $\lambda_{6}$, $\lambda _{7}$, and $\lambda_{8}$ are plotted on the complex $\mathop{\rm Re}(\lambda) $-$\mathop{\rm Im}(\lambda)$ plane (in unit of $\kappa$). The control parameters are varied along the circle shown in the main text. The solid lines denote the trajectories of the calculated eigenvalues.}
	\label{fig_Lep3}
\end{figure*}

\section{Trajectories of the measured Liouvillian eigenvalues associated with the LEP3}
The problem remains of whether or not the other eigenvalues can also display topological features. Neither $\mathbf{V}_0$ nor $\mathbf{V}_8$ coalesces with any other eigenvector, and thus they have nothing to do with the EPs. Consequently, the corresponding eigenvalues $\lambda_0$ and $\lambda_8$ do not exhibit any topological effect. On the other hand, $\mathbf{V}_5$, $\mathbf{V}_6$, and $\mathbf{V}_7$ coalesce at the parameter-space point {$\Omega = \kappa/2$; $\Delta = 0$}. For an HEP3, the spectral winding number defined by Eq.~(8) in the main text is $\pm 1/3$ \cite{Nat.Commun.16.7478}. However, this is not the case for the non-Markovian LEP3. To illustrate this point, we present $\lambda_5$, $\lambda_6$, and $\lambda_7$ in Fig.~\ref{fig_Lep3}, which shows that these eigenvalues move back and forth along the real or imaginary axis, without traversing any closed path, when $k$ is varied from $0$ to $2\pi$. Consequently, the winding numbers associated with these eigenvalues are also zero. The topology associated with an EP3 can also be quantified by the homotopy invariant, calculated in terms of the resultant vector \cite{Nat.Commun.15.10293}.

\begin{figure*}
    \centering
    \includegraphics[width=1.0\textwidth]{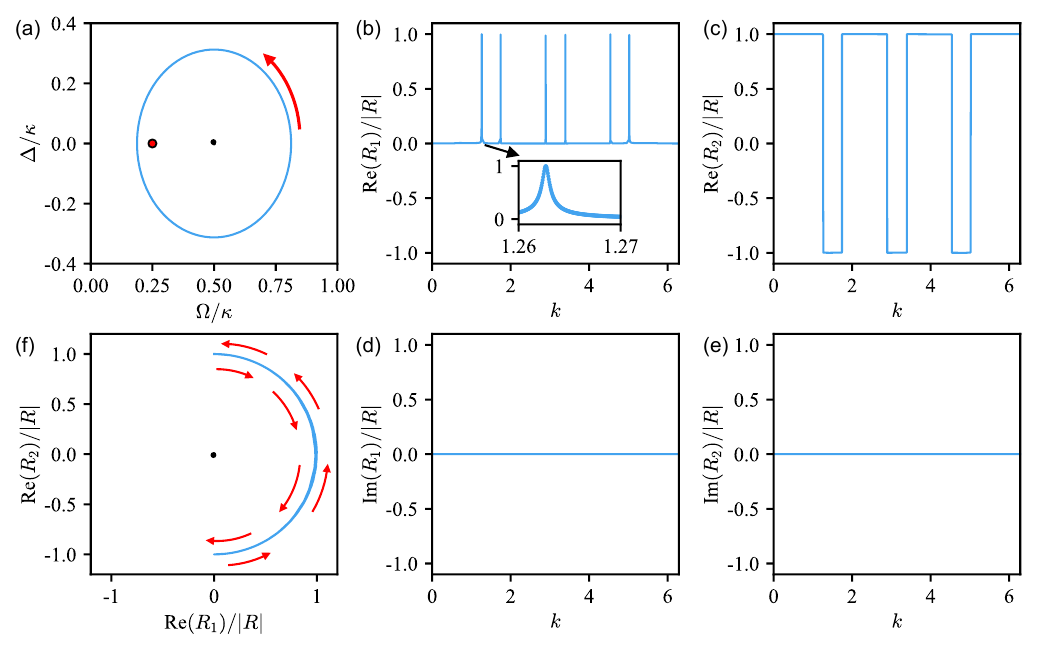}
	\caption{(a) Parameter-space loop. We here set $\Omega=\kappa/2 + r\cos k$  and $\Delta=r\sin k$, so that the LEPs (dot), located at $\{r=\kappa/4, k=\pi\}$, are enclosed in the loop when $r>\kappa/4$. (b)-(e) The components of the resultant vectors with respect to $k$. (f) Trajectory of the rescaled resultant vector on the plane of [Re($R_1$)/$\vert R \vert$, Re($R_2$)/$\vert R \vert$].}
    \label{fig_resultant}
\end{figure*}

The resultant vector $\mathbf{R} = (R_1, R_2)$ is defined by the resultant, 
\begin{equation}
    R_{P_1,P_2} = {\rm det}S_{P_1,P_2},
\end{equation}
where $S_{P_1,P_2}$ is the Sylvester matrix, and $P_i$ can be obtained by the characteristic polynomical $P$,
\begin{equation}
    P_i = \frac{\partial}{\partial \lambda}P,
\end{equation}
where $P = -\prod_{i=1}^{9}(\lambda-\lambda_i)$. The components of the resultant vector are defined as 
\begin{equation}
    \begin{split}
        R_1 =& R_{P_0,P_1},\\
        R_2 =& R_{P_0, P_2}.
    \end{split}
\end{equation}
For an HEP3, such an invariant has a value of ±1, as has been theoretically predicted \cite{Phys.Rev.Lett.127.186602} and experimentally demonstrated \cite{Nat.Commun.15.10293} To obtain the homotopy invariant of LEP3s, we numerically calculate the resultant vector along a closed path in the parameter space, as shown in Fig.~\ref{fig_resultant}(a). The components of the resultant vector as a function of $k$ are displayed in Fig.~\ref{fig_resultant}(b)-(e). It can be seen that the resultant vector just makes a trivial back-and-forth movement even when the control parameter is varied along a closed path, as shown in Fig.~\ref{fig_resultant}(f). This means that the homotopy invariant for the non-Markovian LEP3 is zero.

